\documentclass[preprint2]{aastex}  

\usepackage{graphicx}
\usepackage{amsmath}
\usepackage{natbib}
\usepackage{mathtools}
\usepackage{fancyvrb}
\usepackage{color}
\usepackage{txfonts}
\providecommand{\abs}[1]{\left|#1\right|}
\newcommand{\sdoaia}{{\it SDO}/AIA\ }
\newcommand{\sohoeit}{{\it SOHO}/EIT\ }

\newcommand{\muhz}{$\mu$Hz}

\slugcomment{}

\shorttitle{}
\shortauthors{Auch\`ere et al.}

\begin{document}

   \title{ON THE FOURIER AND WAVELET ANALSIS OF CORONAL TIME SERIES}

	\author{F. Auch\`ere\altaffilmark{1}, C. Froment\altaffilmark{1}, K. Bocchialini\altaffilmark{1}, E. Buchlin\altaffilmark{1} and J. Solomon\altaffilmark{1}}
\affil{Institut d'Astrophysique Spatiale, CNRS, Univ. Paris-Sud, Universit\'e Paris-Saclay, B\^at. 121, 91405 Orsay, France}
   \email{frederic.auchere@ias.u-psud.fr}

\date{\it \center{Received 2016 February 11; revised 2016 April 13; accepted 2016 May 2}}

   \begin{abstract}   
Using Fourier and wavelet analysis, we critically re-assess the significance of our detection of periodic pulsations in coronal loops. We show that the proper identification of the frequency dependence and statistical properties of the different components of the power spectra provies a strong argument against the common practice of data detrending, which tends to produce spurious detections around the cut-off frequency of the filter. In addition, the white and red noise models built into the widely used wavelet code of Torrence \& Compo cannot, in most cases, adequately represent the power spectra of coronal time series, thus also possibly causing false positives. Both effects suggest that several reports of periodic phenomena should be re-examined. The Torrence \& Compo code nonetheless effectively computes rigorous confidence levels if provided with pertinent models of mean power spectra, and we describe the appropriate manner in which to call its core routines. We recall the meaning of the default confidence levels output from the code, and we propose new Monte-Carlo-derived levels that take into account the total number of degrees of freedom in the wavelet spectra. These improvements allow us to confirm that the power peaks that we detected have a very low probability of being caused by noise.
     \end{abstract}
  
%

\section{MOTIVATION}

Following the early detection of an individual case by \cite{Foullon2004}\footnote{Even though these authors attribute this event to a filament, our re-analysis indicates that the locations of significant Fourier power correspond to coronal arcades.},  \cite{Auchere2014} presented evidence that coronal active region loops frequently undergo episodes of periodic pulsations with periods of several hours and lasting several days. This statistical study was based on 13 years of quasi-continuous observations at a cadence of 12 min in the 19.5 nm channel of the {\it Solar and Heliospheric Observatory} \citep[{\it SOHO},][]{Domingo1995} Extreme-ultraviolet Imaging Telescope \citep[EIT;][]{Delaboudiniere1995}. Recently, \cite{Froment2015} published a detailed analysis of three of these pulsation events observed simultaneously in the six coronal passbands of the {\it Solar Dynamics Observatory} \citep[{\it SDO},][]{Pesnell2012} Atmospheric Imaging Assembly \citep[AIA;][]{Lemen2012}. Using a combination of the time-lag analysis method of \cite{Viall2012} and of the differential emission measure (DEM) diagnostics developed by \cite{Guennou2012a, Guennou2012b, Guennou2013}, Froment et al. argued that the observed pulsations are the signatures of incomplete evaporation--condensation cycles similar to those arising in numerical simulations of highly stratified heating of coronal loops \citep{Lionello2013, Mikic2013, Winebarger2014, Lionello2016}. 

Our detections add to an already vast bibliography indicating that periodic phenomena in general seem ubiquitous in the solar corona. In fact, they have been reported to occur in most coronal structures: bright points \citep{Ugarte-urra2004, Tian2008}, loops \citep{Aschwanden1999, Nakariakov1999, Schrijver2002}, prominences \citep{Pouget2006, Bocchialini2011, Zhang2012}, polar plumes \citep{Deforest1998, Ofman1999}, flares \citep{Inglis2008, Nakariakov2009, Dolla2012}, open and closed field large-scale structures \citep{Telloni2013}, etc. The physical processes involved are diverse and the interest of their study is manifold. For example, whether found to be sufficient \citep{McIntosh2011} or insufficient \citep{Tomczyk2007} to compensate for the coronal losses, the energy carried by Alfv\'en waves is at the heart of the long-standing coronal heating debate, and their detection is thus critical. Also, coronal seismology techniques use the characteristics of wave and oscillatory phenomena to derive the properties of the ambient plasma and magnetic field \citep{Nakariakov2007, DeMoortel2012}.

However, recent papers have questioned the validity of several accounts of quasi-periodic pulsations (QPPs) in solar flares \citep{Gruber2011, Inglis2015, Ireland2015}. These authors point out that, in many cases, the fundamental power-law dependence of the power spectra of coronal time series has not been recognized, resulting in erroneous confidence levels. After the re-analysis of 19 time series, \cite{Inglis2015} concluded that coherent oscillatory power is necessary to explain the observed Fourier spectra in only one case. Beyond the specific case of QPPs, their results cast doubt on a number of previously published reports and suggest that the prevalence of oscillatory phenomena in the corona may be artificial.

In this context, even though we properly accounted for the power-law nature of the coronal power spectra in our previous works, we decided to critically re-evaluate the statistical significance of the events presented by \cite{Froment2015}. Indeed, several sometimes subtle effects can cause false positives in Fourier or wavelet analyses. For example, Appendix A of \cite{Auchere2014} describes how the shadow cast on the detector by the mesh grid holding the focal filters in \sohoeit \citep{Defise1999, Auchere2011} produces spurious frequencies in the range of those detected in coronal loops. This effect could not be found in identical processing of AIA images, which is explained by the much smaller amplitude of the mesh grid pattern in this latter instrument. In the present paper, before re-interpreting our previous detections, we present possible sources of false positives in the methods of spectral (Fourier and wavelet) analysis: section~\ref{sec:detrending} describes the effect of time series detrending on the power spectra, section~\ref{sec:noise_model} discusses the choice of noise model, and section~\ref{sec:confidence_levels} treats confidence levels in wavelet spectra. As an example, one of the periodic pulsation events studied by \cite{Froment2015} is then re-analyzed in section~\ref{sec:aia_application} in the light of these considerations. Our conclusions and recommendations for Fourier and wavelet analysis of coronal time series are summarized in section~\ref{sec:summary}.



\section{DETRENDING\label{sec:detrending}}
\begin{figure*}[t]
	\centering
		\includegraphics[width=\textwidth]{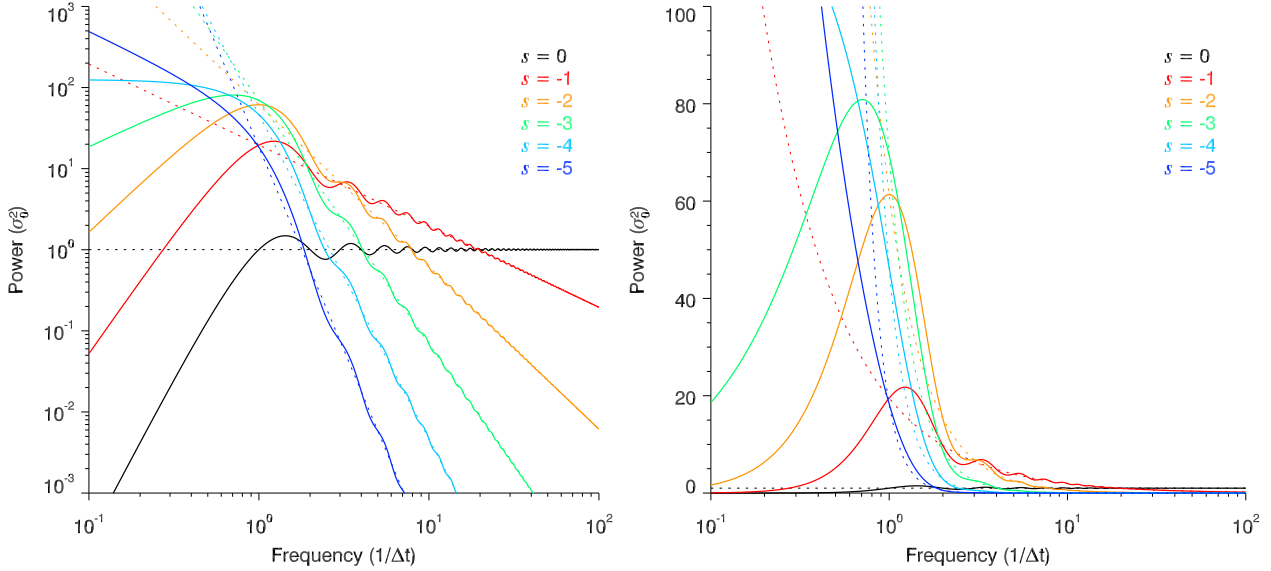}
		\caption{Distortion of the power spectra of time series after detrending with a running boxcar of width $\Delta t$. Left: the original time series have power-law-like spectra (dashed lines for power-law exponents $s$ varying from 0 to -5). The abscissas are expressed in units of the cut-off frequency $1/\Delta t$ of the filter and each curve is normalized to its average over the plotting range, i.e. to the variance $\sigma_0^2$ of the corresponding time series. For $s>-4$, the power spectra of the detrended time series (solid lines) have a maximum around the cut-off frequency, which results from the suppression of the low-end of the power-laws by the high-pass filter (black curve, $s=0$). Right: the same as on the left but with a linear axis of power. The power spectra of the detrended series have dominant power in a limited band of frequencies, which can result in a convincing but false impression of periodicity in the original signal (see Figures~\ref{fig:detrended_wavelet} and~\ref{fig:wrong_spectrum}).}
	\label{fig:detrended_power_spectrum}
\end{figure*}

\begin{figure*}[htbp!]
 \centering
		\includegraphics[width=\textwidth]{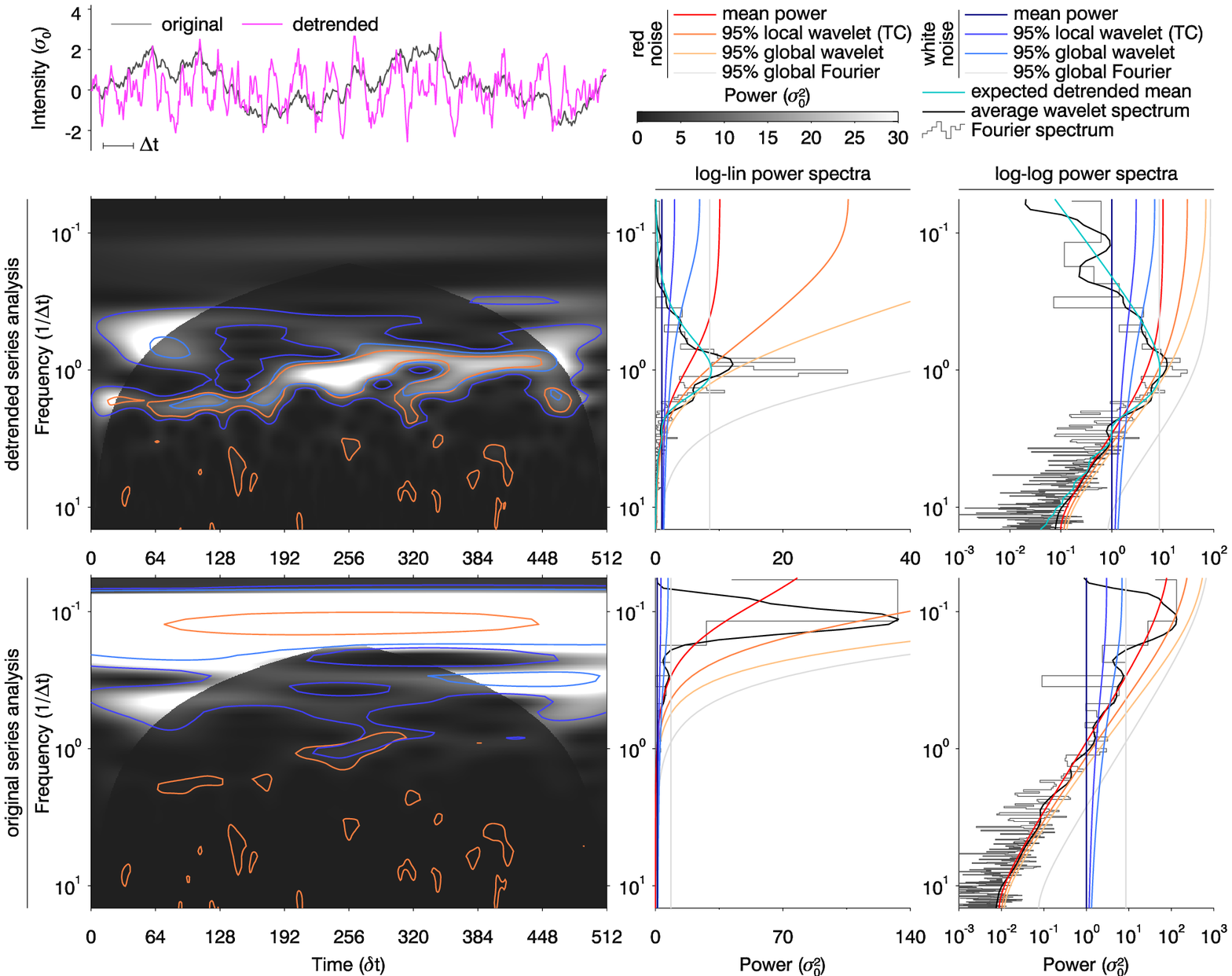}
		\caption{Spurious periodicity produced by detrending a random time series. Top left: the original simulated data (in gray) have a mean power spectrum that is a power-law of exponent -2 (histogram of the bottom right panel). Its detrended version (with a running boxcar of width $\Delta t=30$) is plotted in magenta. The wavelet and Fourier analyses of the detrended and original data are shown in the middle and bottom rows respectively. The clear $\Delta t$ periodicity visible in the detrended time series manifests itself as a narrow band of excess power in the wavelet spectrum and as a strong peak near the cut-off frequency $1/\Delta t$ in the Fourier and time-averaged wavelet spectra (gray histograms and black curves). However, this shape is simply what is expected from the  high-pass filtering of a power-law spectrum (cyan curve, i.e. orange curves of Fig.~\ref{fig:detrended_power_spectrum}), as best revealed by a log--log representation (right panels). The COI, i.e. the region of the wavelet spectrum affected by the zero padding of the time series, is shown in lighter shades of gray.}
	\label{fig:detrended_wavelet}
\end{figure*}

Detrending is often applied to time series before analysis of their frequency content in order to filter out the low frequencies and thus to enhance the periodic signals potentially present in the original data \citep[e.g.][]{Foullon2004, Ugarte-urra2004, Inglis2008, Tian2008, Foullon2009, Nakariakov2010, Dolla2012}. However, several authors \citep{Cenko2010, Iwakiri2010, Gruber2011} have shown that detrending, combined with an improper accounting for the frequency dependence of the power spectrum, can lead to overestimating the claimed significance levels or even to false detections. Our analysis confirms and generalizes these results. We demonstrate that detrending not only {\it can} but {\it generally does} lead to false-detections in Fourier or wavelet analyses, although it produces visually convincing filtered versions of the time series.


Detrending is achieved by removing from the original time series $s_o(t)$ a smooth version of it, which is commonly obtained with high-pass filtering.\footnote{Other methods can be used -- like approximation by polynomials -- but in all cases the power spectrum will be affected.} Some authors compute the detrended time series $s_d(t)$ by subtracting the smooth estimate from the original data \citep[e.g.][]{Ugarte-urra2004, Inglis2008, Nakariakov2010, Dolla2012}
\begin{equation}
s_d(t) = s_o(t) - s_o(t) * f(t),
\label{eq:detrend_subt}
\end{equation}

\noindent
where $s_o(t) * f(t)$ represents the smoothed data obtained by convolution of $s_o(t)$ with the filter kernel $f(t)$. Other authors divide the original data by the smooth estimate and subtract 1 \citep[e.g.][]{Fludra2001, Foullon2004, Foullon2009} to obtain
\begin{equation}
s_d(t) = \frac{s_o(t)}{s_o(t) * f(t)} - 1.
\label{eq:detrend_mult}
\end{equation}

\noindent
If the variations $\delta s_o(t)$ of the signal are small with respect to its global average $\overline{s_o}$, equation~\ref{eq:detrend_mult} can be expressed as 
\begin{equation}
s_d(t) \approx \frac{\delta s_o(t) - \delta s_o(t) * f(t)}{\overline{s_o(t)}},
\end{equation}
 
\noindent
which is identical to equation~\ref{eq:detrend_subt} for the relative variations of the signal. These two methods are thus equivalent and have the same effect on the Fourier or wavelet spectra. If the variations of the signal are not small compared to its average, the second method modifies the power spectra more profoundly than described below, which renders their proper interpretation impossible.


Using the convolution theorem on Equation~\ref{eq:detrend_subt}, we obtain the power spectrum of the detrended time series
\begin{equation}
\Psi(\nu)=\abs{S_d(\nu)}^2=\abs{S_o(\nu)}^2\abs{1-F(\nu)}^2,
\label{eq:generic_detrended_power_spectra}
\end{equation}
\noindent
where capital letters are used to denote the Fourier transforms. The most commonly used high-pass filter is a running boxcar of width $\Delta t$, in which case the power spectrum becomes\footnote{Similar expressions could be derived for other high-pass filters, resulting in similar distortions of the power spectrum.}
\begin{equation}
\Psi(\nu)=\abs{S_o(\nu)}^2\left(1 - \frac{\sin\left(\pi\nu\Delta t\right)}{\pi\nu\Delta t}\right)^2,
\label{eq:boxcar_detrended_power_spectra}
\end{equation}
\noindent
which is the power spectrum of the original time series multiplied by a filtering function that attenuates the low frequencies while preserving the high frequencies (the black curve in Figure~\ref{fig:detrended_power_spectrum}).

As recently emphasized by \cite{Inglis2015} and \cite{Ireland2015}, the mean Fourier power spectra of extreme-ultraviolet emission from active region cores, loop footpoints, and the quiet Sun follow a power-law-like behaviour. This property was already noted by \cite{Auchere2014} for all types of on-disk structures and for most of the time series that they analyzed, and \cite{Gupta2014} also reports power-law spectra off-disk in polar plumes. A power-law is actually what is expected from line of sight superimposition of many exponentially decaying emission pulses \citep{Ireland2015, Aschwanden2016}, which is consistent with the idea that the corona is heated by many small scale impulsive events. This indicates that a power-law is the most likely spectral shape for coronal time series. From Equation~\ref{eq:boxcar_detrended_power_spectra}, the expected power spectra of detrended time series can thus be described by
\begin{equation}
\Psi(\nu) = a\nu^s\left(1 - \frac{\sin\left(\pi\nu\Delta t\right)}{\pi\nu\Delta t}\right)^2.
\end{equation}
\noindent
$\Psi(\nu)$ is represented in Figure~\ref{fig:detrended_power_spectrum} for integer values of the exponent $s$ from -1 to -5,  which covers the range of observed slopes reported by \cite{Inglis2015}. In addition, the shape of the filter itself is given by the  $s=0$ case. 
For $s\leqslant -4$, the maximum of $\Psi(\nu)$ is at $\nu=0$. For $-1>s>-4$, which is the case for the majority of the reported power-laws for coronal time series, we determined numerically that the position of the maximum can be approximated (to within better than 1\%) by
\begin{equation}
\frac{1}{\Delta t}\sqrt{\frac{4+s}{2}}.
\end{equation}
\noindent
The maximum is located near the cut-off frequency $1/\Delta t$ of the high-pass filter, and exactly at the cut-off for $s=-2$. The right panel of Figure~\ref{fig:detrended_power_spectrum}, which uses a linear axis for the power, clearly illustrates the formation of a peak of dominant frequencies around the cut-off by the filtering out of the power-laws below $1/\Delta t$. The corresponding detrended time series are thus strongly chromatic, which can be incorrectly interpreted as periodicity in the original data.

In order to demonstrate the effect in practice, using an independently derived version of the algorithm described by \cite{Timmer1995}, we simulated a random time series of $N=512$ data points whose power spectral distribution (PSD) is a power-law of exponent -2. The original time series is the gray curve in the top left panel of Figure~\ref{fig:detrended_wavelet} and its $\Delta t=30$ time steps wide running boxcar detrended version is in magenta. Both curves are normalized to their respective standard deviations $\sigma_0$. The results of the spectral analysis (wavelet and Fourier) of the detrended and original series are represented respectively in the middle and bottom rows, respectively.

A periodicity is clearly visible in the detrended time series even though the original data are completely random. This periodicity is real, as shown by the narrow peak of power in the log--linear fast Fourier transform power spectrum of the central panel. A narrow band of power is also visible at the same frequencies in the Morlet wavelet spectrum (left panel) computed with the code of \citet[][hereafter TC98]{Torrence1998}.

Fourier (gray line) and wavelet (blue contours and lines) white noise 95\% confidence levels could be used to argue in favor of the significance of the peak. However, as revealed by the log--log representation of the right panel, this periodicity is only due to the high-pass filtering of a power-law-like spectrum. The Fourier (gray histogram) and time-averaged wavelet (black curve) power spectra follow the theoretical curve given by Equation~\ref{eq:boxcar_detrended_power_spectra} (in cyan, i.e. the $s=-2$ curve of Figure~\ref{fig:detrended_power_spectrum}). As expected, the peak of power is located around the cut-off frequency $1/\Delta t$. At higher frequencies, the power spectrum is quasi-unaffected and resembles that of the original series (lower right panel). As we will see in the next section, not recognizing the power-law nature of the original spectrum will lead to incorrect conclusions regarding the significance of the observed peak of power.

\section{BACKGROUND NOISE MODELS\label{sec:noise_model}}

\begin{figure*}[htbp!]
	\centering
		\includegraphics[width=\textwidth]{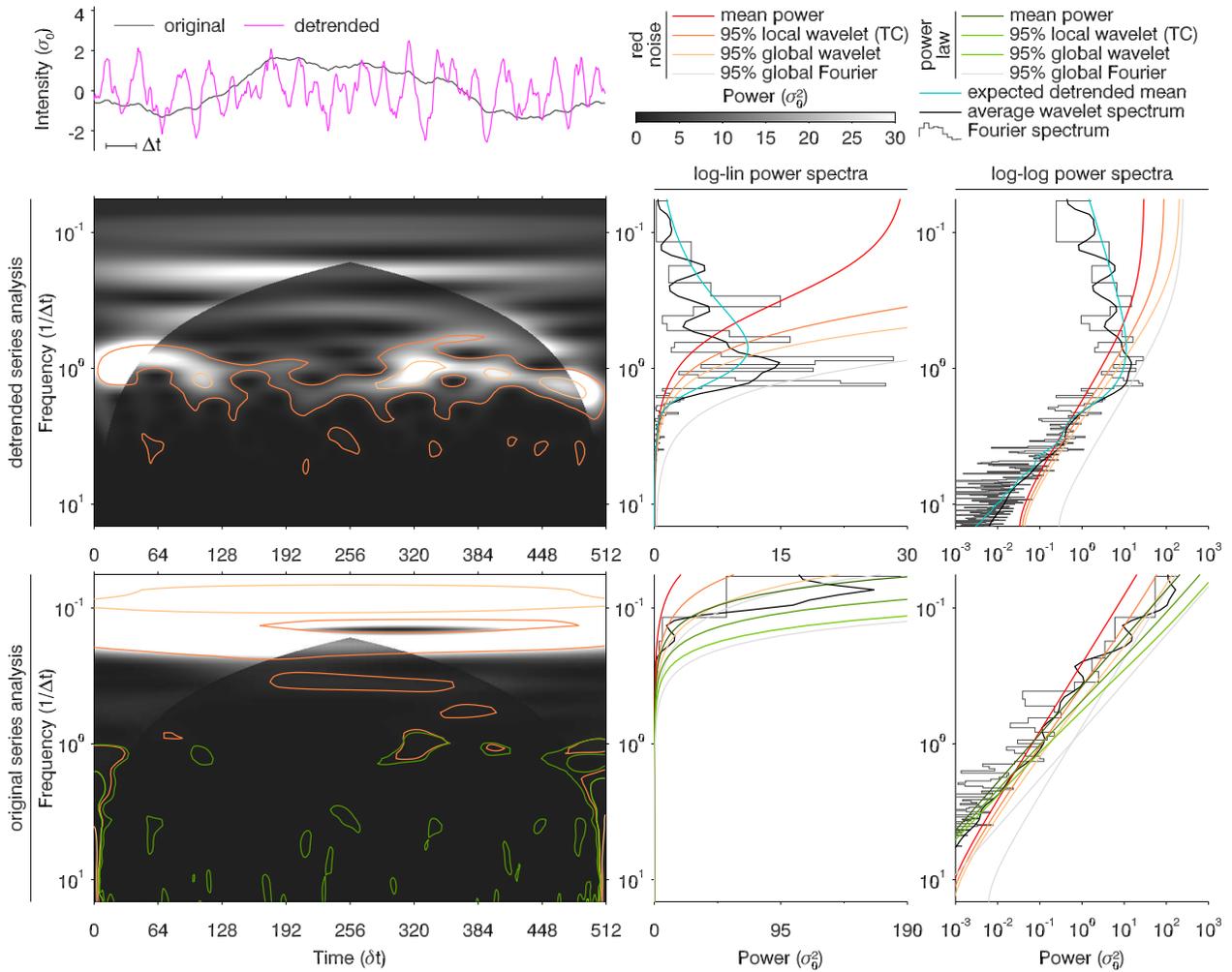}
		\caption{Similar to Figure~\ref{fig:detrended_wavelet} for a random time series whose mean power spectrum is a power-law of exponent -3. The AR(1) model (red curves) is not a valid approximation of the spectrum of the detrended (middle row) or of the original (bottom row) time series, yielding erroneous confidence levels and conclusions. Confidence levels based on a power-law model of the expected spectrum (dark green curves) allow us to correctly conclude the absence of significant oscillatory power.}
	\label{fig:wrong_spectrum}
\end{figure*}

The determination of confidence levels requires an appropriate model of the background power, i.e. an estimate of the expected value $\sigma(\nu)$ of the power, and of its probability distribution, at each frequency in the absence of oscillatory phenomena.\footnote{As mentioned by \cite{Gabriel2002}, the commonly adopted notation $\sigma$ for the expected power is not strictly correct, and it is not to be confused with the variance $\sigma_0^2$ of the time series, even though the two are equal for white noise.} This is a fundamental step in wavelet or Fourier analysis\footnote{It is worth noting that randomization methods \citep{Oshea2001} do not replace the assumption of a noise model since they are in fact equivalent to comparing to white noise.}. If we assume white noise, the expected power $\sigma$ is constant (dark blue lines in the two right-most panels of the middle row of Figure~\ref{fig:detrended_wavelet}) and equal to 1 since, following TC98, we normalized the Fourier and wavelet spectra by $N/\sigma_0^2$, which gives a measure of power relative to white noise. For a normally distributed random variable, the Fourier power at each frequency being distributed as $\sigma\chi_2^2/2$ \citep[TC98,][]{Gabriel2002}\footnote{The notation $\sigma\chi_d^2/d$, as used by TC98, means that the probability distribution function (PDF) of the power $p$  at the frequency $\nu$ is $f\left(p\ d/\sigma;d\right)d/\sigma$, where $f(x;d)$ is the PDF of the chi-square distribution of degree $d$. This distribution of power has mean $\sigma$.}, the probability that the power is greater than $m$ times the mean $\sigma$ for at least one of {\it any} of the $N/2$ frequencies is given by $1-\left(1-\text{e}^{-m}\right)^{N/2}$ \citep{Scargle1982, Gabriel2002}, from which we obtain that the peaks of power above $m\sigma = -\log\left(1 - 0.95^{1/256}\right)\sigma= 8.51\sigma$ have only a 5\% chance of occurring from white noise (95\% {\it global} confidence levels, the gray lines in the two right-most panels).

Equivalently, we derived empirically (see~\S~\ref{sec:confidence_levels}) the 95\% {\it global} confidence levels above which white noise power only has a 5\% probability to lie for at least one of {\it any} of the points of the wavelet and time-averaged wavelet spectra (light blue contours and curves). The less stringent {\it local} confidence levels built-in the TC98 code (medium blue contours and curves) give the 95\% probability that greater power at {\it each} frequency and/or time is not due to white noise. Although perfectly rigorous for white noise, these confidence levels lead to erroneous conclusions because this background model is obviously inadequate for either the original or the detrended time series.


The red noise model of \citet{Torrence1998} (red curves) is not a good approximation of the expected spectrum of the detrended series either. It is based on the lag-1 auto-regressive (AR(1), or Markov) process:
\begin{equation}
x_n=\alpha x_{n-1}+z_n,
\label{eq:ar1_process}
\end{equation}
\noindent
where $\alpha$ is the lag-1 autocorrelation, $x_0=0$ and $z_n$ is taken from Gaussian white noise. The normalized expected Fourier power spectrum of the resulting time series is given by:
\begin{equation}
\sigma(\nu)=\frac{1-\alpha^2}{1+\alpha^2-2\alpha\cos\left(2\pi\nu\delta t\right)},
\label{eq:ar1_process_power}
\end{equation}
\noindent
where $\delta t$ is the sampling interval. Since $x_n$ is a normally distributed random variable, its power is distributed as white noise of mean $\sigma(\nu)$ in each sufficiently narrow frequency band centered on $\nu$. Thus, the Fourier and wavelet spectra are distributed as $\sigma(\nu)\chi_2^2/2$ at every frequency and time\footnote{This expression is valid for complex wavelets such as the Morlet and Paul wavelets used in the present paper.} (see TC98 and references therein) and the corresponding confidence levels are simply those derived for white noise multiplied by $\sigma(\nu)$.

As was the case for white noise, using the red noise AR(1) model to study the power spectrum of the detrended time series of Figure~\ref{fig:detrended_wavelet} naturally yields misleading results because the model cannot represent the expected spectrum. The detrended time series has a lag-1 autocorrelation coefficient of 0.83. The corresponding mean power (red curves in the right panels of the middle row) underestimates the expected value (cyan curves) around the cut-off $1/\Delta t$ and largely overestimates it at lower frequencies. Consequently, while the Fourier power is everywhere below the 95\% confidence level (gray curves), both the wavelet and the time-averaged wavelet spectra exceed the TC98 95\% {\it local} confidence levels (orange contours and curves).

Conversely, the AR(1) model can adequately represent the mean power of the non-detrended time series (red curves in the bottom right panels) since this latter has a lag-1 autocorrelation coefficient of 0.995. Indeed, for $\alpha\approx 1$ and small $\nu$, using a second order Taylor expansion of the cosine function, Equation~\ref{eq:ar1_process_power} can be approximated by
\begin{equation}
\sigma(\nu)\approx\frac{1-\alpha}{2\left(\pi\nu\delta t\right)^2},
\label{eq:approx_ar1_power}
\end{equation}
\noindent
which is a power-law of exponent -2, i.e. by construction the expected PSD of the original data.

As a result of the appropriateness of the model, none of the bins of the Fourier and time-averaged wavelet spectra are above the associated 95\% {\it global} confidence levels (gray and yellow curves respectively). However, at least one point of the time-averaged wavelet spectrum is above the TC98 95\% {\it local} confidence level (orange curves), but this should not be taken as evidence for significant oscillatory power, for it is in fact likely to occur by chance: while the probability for the noise power at {\it each} frequency to be above that level is only 5\%, the probability for {\it any} of the $256$ frequencies to be above is much higher (see \S~\ref{sec:confidence_levels}). Likewise, several orange contours (TC98 95\% {\it local} confidence levels) appear by chance on the wavelet spectrum (bottom left panel). This effect was mentioned in TC98 (see their Fig. 4) and the authors caution against over-interpretation on their website.\footnote{\url{http://paos.colorado.edu/research/wavelets/faq.html}} On the other hand, no points of the wavelet or time-averaged spectra are above the 95\% {\it global} confidence levels derived in \S~\ref{sec:confidence_levels} (yellow contours and curves).

The AR(1) red noise model is nonetheless strongly limited in that, as shown by Equation~\ref{eq:approx_ar1_power}, it can only satisfactorily approximate power-law-like spectra of exponent -2, while the values reported in the literature for coronal time series span at least the -1.72 to -4.95 range \citep{Inglis2015, Ireland2015}. It is thus likely that the AR(1) model is in most cases {\it not} pertinent for solar data.

Figure~\ref{fig:wrong_spectrum} is similar to Figure~\ref{fig:detrended_wavelet} for a random time series having an expected power-law spectrum of exponent -3. The detrended time series (in magenta) presents strong oscillations, for the same reasons as described above. Its spectral analysis (middle row) thus presents characteristics similar to those illustrated in Figure~\ref{fig:detrended_wavelet} for the $s=-2$ series. The false detections appear to be even more significant because in this case the AR(1) noise model (with $\alpha=0.92$) is not a valid approximation of the PSD anywhere. The non-detrended data have a lag-1 autocorrelation coefficient of 0.998. The mean power spectrum given by the AR(1) model is thus close to a power-law of exponent -2. It intersects the true PSD at mid-frequencies, and the corresponding 95\% confidence levels erroneously lead to the conclusion that there is significant excess power at low frequencies.

Supposing that we ignore how the time series was built, the log--log representation of its PSD (bottom right panel) would naturally suggest choosing a power-law of the variable exponent as a noise model. The thus fitted expected power is given by the dark green line. The Fourier spectrum is not above the corresponding 95\% {\it global} confidence level (gray line) anywhere. Wavelet confidence levels can in fact be computed with the TC98 code for any given mean power spectrum (see the Appendix for practical details). Some points of the wavelet and time-averaged wavelet spectra are above the resulting {\it local} 95\% confidence levels (medium green contours and curves) for the same reason as described above: these levels account only for the local number of degrees of freedom of the spectra, and thus power will randomly surpass them for 5\% of the frequencies and/or times even if the noise model is appropriate. In the next Section, we derive {\it global} confidence levels that account for this effect.

\section{CONFIDENCE LEVELS\label{sec:confidence_levels}}

 \subsection{Fourier Confidence Levels\label{sec:fourier_levels}}

At each frequency $\nu$, the Fourier or (wavelet) power spectrum of a random time series is distributed as \citep[][TC98]{Gabriel2002}\footnote{This expression is valid for complex wavelets. The 1/2 factor would be removed for real wavelets (see TC98)}
\begin{equation}
\frac{1}{2}\sigma(\nu)\chi_2^2
\label{eq:fourier_wavelet_local_distribution}
\end{equation}
\noindent
around the mean power $\sigma(\nu)$. The probability for one point of a spectrum to have a power greater than $m$ times the mean $\sigma$ is thus $P(m)=e^{-m}$, and the associated confidence level as defined by TC98 is $1-P$. For example, points above the TC98 95\% confidence level have a probability $P=0.05$ (5\%) of having a power greater than $m=-\log(0.05)\sigma(\nu)=2.99\sigma(\nu)$. 

However, in most practical situations, the probability of having power greater than these confidence levels in at least one of the bins is close to one. The reason is that the confidence levels above which peaks of power have a given probability to occur by chance depend on the number of points in the spectrum. For a random time series of $N$ points, $1-P(m)$ is the probability for each of the $N/2$ frequency bins of the Fourier spectrum to have a power lower than $m\sigma(\nu)$. Since the bins are independent, $(1-P(m))^{N/2}$ is the probability for the power to be lower than $m\sigma(\nu)$ in all bins, and the probability that at least one bin has a power greater than $m\sigma(\nu)$ is thus
\begin{equation}
P_g(m)=1 - (1 - P(m))^{N/2} = 1 - (1 - e^{-m})^{N/2}.
\label{eq:fourier_global_probability}
\end{equation}
\noindent
We refer to $P_g$ as the {\it global} probability (and corresponding confidence levels) over the whole spectrum as opposed to the {\it local} probability associated with individual bins.\footnote{The term {\it global} refers to the taking into account of the total number of degrees of freedom in the spectrum and is not to be confused with the {\it global wavelet spectrum} term used in TC98. To avoid confusion, in this paper we always use {\it time-averaged spectrum} to denote what TC98 also call the {\it global wavelet spectrum}.} For example, for $N=512$, the {\it global} probability to have at least one bin of the Fourier spectrum above the 95\% {\it local} confidence level is $1 - (1 - 0.05)^{256}=0.999998$. Conversely, using Equation~\ref{eq:fourier_global_probability}, one can properly derive the value of $m$ corresponding to the 95\% {\it global} confidence level, i.e. $m=-\log(1-0.95^{1/256})= 8.51$.

\subsection{Wavelet Confidence Levels\label{sec:specificities}}

\subsubsection{How Significant are Peaks of Wavelet Power?\label{sec:specificities}}

\begin{figure*}[htbp!]
\centering
		\includegraphics[width=0.75\textwidth]{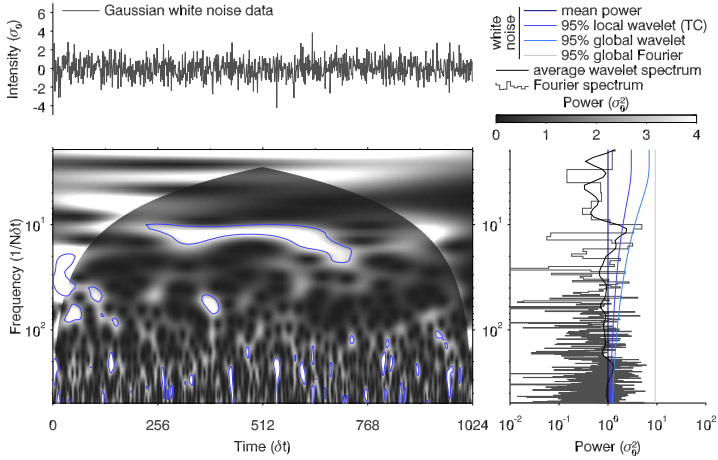}
		\caption{Example of coherent structure in the Morlet wavelet power spectrum of a Gaussian white noise time series. Around the frequency $12/N\delta t$, the power exceeds the 95\% {\it local} confidence level during about 500~$\delta t$ (dark blue contours). The possibility that such features can occur randomly indicates that comparison of the width of a peak of power with the decorrelation time is not a sufficient test to conclude its significance. The {\it global} confidence levels defined in \S~\ref{sec:confidence_levels} on the other hand allow us to correctly conclude the absence of significant peaks in both the Fourier, wavelet, and time-averaged wavelet spectra.}\label{fig:random_wavelet}
\end{figure*}

\begin{figure}[htbp!]
\centering
		\includegraphics[width=0.456\textwidth]{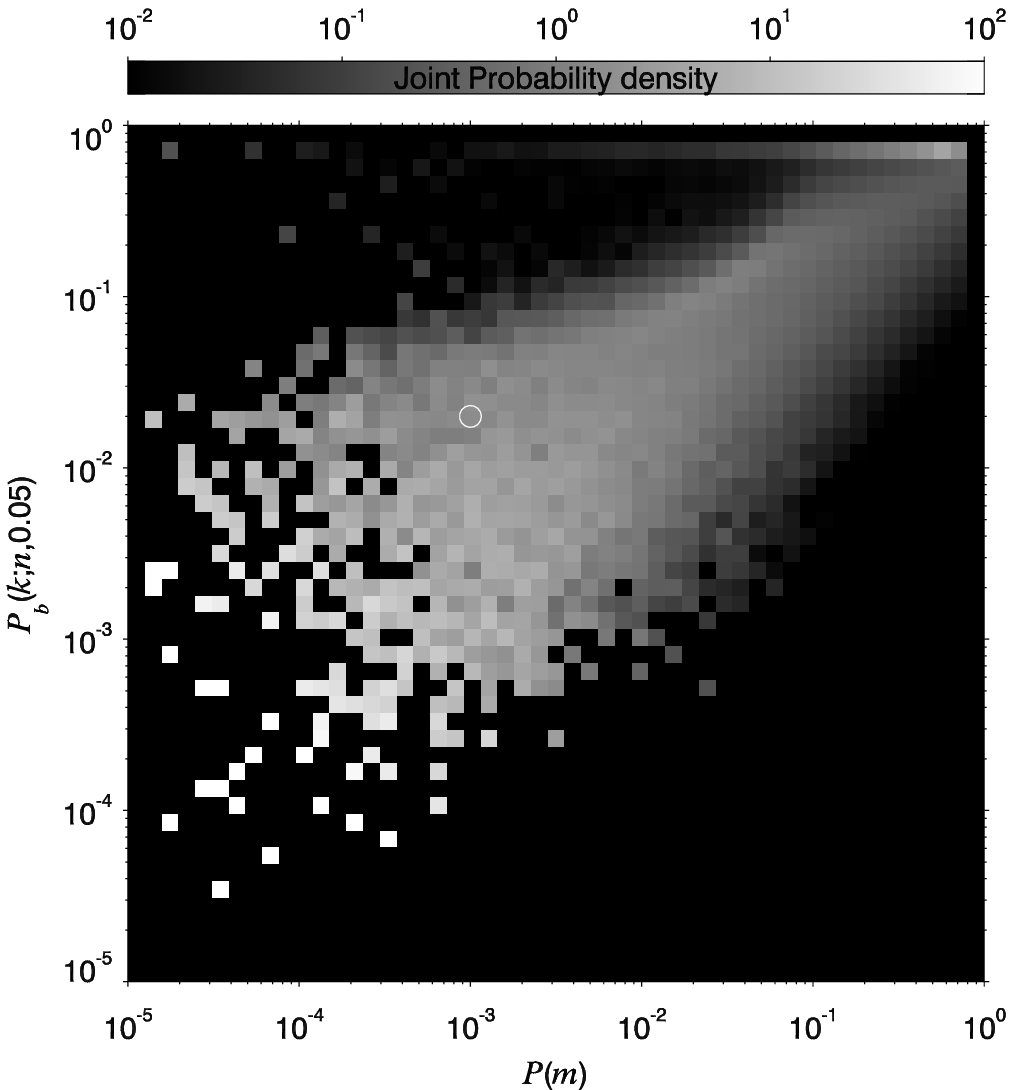}
		\caption{Joint probability density between the probability associated with a power level in a time-averaged spectrum and the probability of having the observed number of bins above the local 95\% confidence level at the corresponding scale. The white circle indicates the corresponding values for the long structure of Figure~\ref{fig:random_wavelet}. The correlation shows that estimating the probability of occurrence of such a structure amounts to estimating the probability associated with the corresponding peak in the time-averaged spectrum.}\label{fig:binomial_correlation}
\end{figure}

The derivation of global confidence levels for a {\it wavelet} spectrum is more complex because its bins are not statistically independent. Without providing a quantitative test, Torrence \& Compo 
suggest that the significance of a peak of power be judged by comparing its duration with the decorrelation time, which is given by the width of the cone of influence (COI) at the corresponding frequency. This approach is followed by, e.g., \cite{Jess2008}. But there is a non-zero probability for two or more random peaks to be close enough together to form structures longer than the decorrelation time. Figure~\ref{fig:random_wavelet} shows an example in the wavelet spectrum of a 1024 data-point-long Gaussian white noise time series. The structure visible around the frequency $12/N\delta t$ remains above the 95\% {\it local} confidence level (dark blue contours) for about 4 times the decorrelation time. As shown below, the probability of occurrence of such a structure at a given frequency (or scale) can be estimated using the binomial distribution.

TC98 showed that the time-averaged wavelet power is distributed as $\sigma(\nu)\chi_d^2/d$, with $d$ being the number of degrees of freedom at each scale $s$
\begin{equation}
d=2\sqrt{1+\left(\frac{n_a\delta t}{\gamma s}\right)^2},
\label{eq:time_averaged_dof}
\end{equation}
\noindent
where $n_a$ is the total number of points $N$ minus half the number of those in the COI, and $\gamma$ is an empirically derived decorrelation factor equal to 2.32 for the Morlet wavelet. At each scale, the wavelet spectrum thus behaves as $l=d/2$ statistically independent $\chi_2^2$ distributed bins. Defining a success as a peak of power surpassing the 95\% local confidence level, which has probability $p=0.05$, the probability of obtaining exactly $k$ successes in $l$ trials is given by
\begin{equation}
P_b(k;l,p)=\frac{l!}{k!(l-k)!}p^k(1-p)^{l-k}.
\label{eq:binomial}
\end{equation}
\noindent
In practice, $l$ and $k$ do not have to be integers; in this case the Gamma function has to be used instead of factorials. We have $l=4.5$ at the scale of the long structure of Figure~\ref{fig:random_wavelet} and $k=1.8$ given that it lasts for 40\% of the total duration. The probability to obtain the same number of bins above the 95\% local confidence level by chance is thus $2\times 10^{-2}$. Since the binomial distribution accounts for all possible arrangements of $k$ successes in $n$ trials, this value is in fact an upper-limit for the probability of occurrence of that particular structure, which may thus be considered unlikely. This structure corresponds to a peak of power of the time-averaged spectrum (the black curve in the right panel) that lies above the 95\% {\it local} confidence level (dark blue curve). Knowing that the time-averaged spectrum is distributed as $\sigma(\nu)\chi_d^2/d$ (TC98), the probability associated with this maximum of power is $10^{-3}$, which is also unlikely.

In fact, it is intuitive that there must be a correlation between the probability associated with a power level in the time-averaged spectrum and the probability to have the observed number of bins above the local 95\% confidence level at the corresponding scale. Figure~\ref{fig:binomial_correlation} shows, for 100,000 wavelet spectra of Gaussian white noise time series, the joint probability density of these two quantities. 
The spread around the diagonal comes from the binomial approximation and from the fact that at a given scale, several temporal power profiles can produce the same time-averaged power while having different number of bins above the 95\% local confidence level. The observed correlation demonstrates that estimating the probability of occurrence of a given coherent structure simply amounts to estimating the probability associated with the corresponding time-averaged power. In order to do this, one must nonetheless take into account the total number of degrees of freedom in the spectrum. In the example of Figure~\ref{fig:random_wavelet}, while the time-averaged wavelet power exceeds the 95\% {\it local} confidence level, the 95\% {\it global} Fourier confidence level given by Equation~\ref{eq:fourier_global_probability} (gray line) correctly rejects the corresponding peaks in the Fourier spectrum (gray histogram). Likewise, it is possible to define robust {\it global} confidence levels for both wavelet and time-averaged wavelet spectra by generalizing the principles described ~\ref{sec:fourier_levels} for Fourier spectra.

\subsubsection{Global wavelet confidence levels\label{sec:global_confidence_levels}}

Equation~\ref{eq:fourier_global_probability}, which was derived for Fourier spectra, cannot be used for wavelet spectra by simply replacing $N/2$ with the total number of bins, for these are not statistically independent. Nonetheless, as shown below, the relation between $P_g(m)$ and $P(m)$ can in practice be approximated by a modified version of Equation~\ref{eq:fourier_global_probability}
\begin{equation}
P_g(m)=1 - (1 - P(m)^a)^n
\label{eq:wavelet_global_probability}
\end{equation}
\noindent
where $a$ and $n$ are empirically derived coefficients. By analogy with Equation~\ref{eq:fourier_global_probability}, $n$ is expected to be proportional to the number of bins in the spectrum, but limited to those outside the COI since the {\it local} confidence levels corresponding to $P(m)$ are not relevant for the bins inside the COI. Indeed, for a zero-padded time series, the mean power inside the COI decreases as $1-e^{-2t/\tau_s}/2$, where $\tau_s$ is the {\it e}-folding time (equal to $\sqrt{2}s$ for the Morlet wavelet) and $t$ is the time from either the beginning or end of the spectrum. Thus, bins inside the COI have a reduced probability of being above the chosen confidence level compared to those outside. Finally, the total number of bins outside the COI should be normalized to the resolution in scale because the probability of having at least one bin above a given confidence level should be independent of the resolution.

In order to determine $a$ and $n$, we constructed 100,000 Gaussian white noise time series\footnote{Note that since, as shown by Equation~\ref{eq:fourier_wavelet_local_distribution}, the power is always distributed as $\chi_2^2$ around the mean power $\sigma(\nu)$, the Monte-Carlo results would be identical for, e.g., AR(1) or power-law noise.} along with their associated Morlet wavelet power spectra, from which we estimated the probability $P_g(m)$ to have at least one bin (outside the COI) above the power threshold $m\sigma(\nu)$ and corresponding to the local probability $P(m)$. We repeated this procedure for different numbers of data points $N$ (powers of 2 from $2^6$ to $2^{14}$) and scale resolutions $\delta j$ (from $1/2$ to $1/2^8$). The wavelet spectra were all computed using zero padding over $J+1$ scales $s_j=s_02^{j\delta j}$ with $J=\log_2(N\delta t/s_0)/\delta j$ and the minimum resolvable scale $s_0=2\delta t$ (see TC98 for details on the meaning of these parameters). The total number of bins in each spectrum is $N(J+1)$.
\begin{figure}
	\centering
		\includegraphics[width=0.456\textwidth]{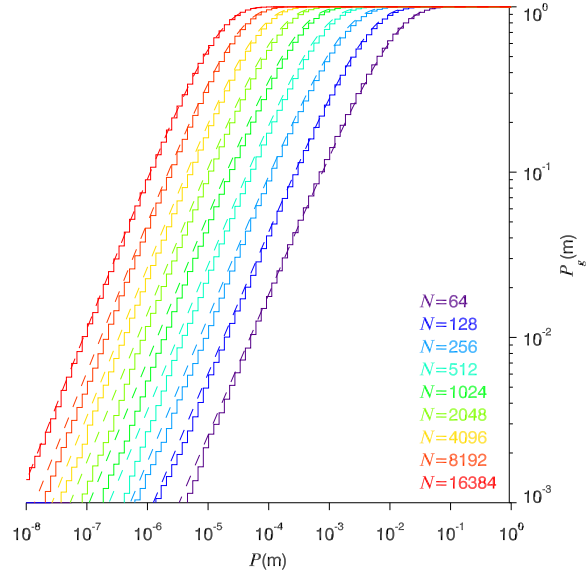}
		\caption{Monte-Carlo-derived probabilities $P_g(m)$ that {\it at least one} bin of a Morlet wavelet spectrum has power greater than $m$ times the mean power $\sigma(\nu)$, as a function of the probability $P(m)$ for {\it each} bin to have a power greater than $m\sigma(\nu)$, for several lengths $N$ of the input random time series. The dashed curves correspond to the parameterization given by Equations~\ref{eq:wavelet_global_probability},~\ref{eq:wavelet_para_a} and~\ref{eq:wavelet_para_w}.}\label{fig:torrence_statistics_spectrum}
\end{figure}

\begin{figure}
	\centering
		\includegraphics[width=0.456\textwidth]{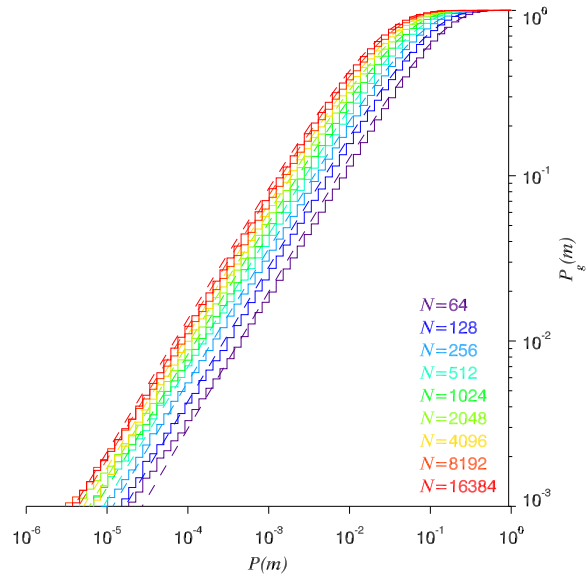}
		\caption{The same as figure~\ref{fig:torrence_statistics_spectrum} for time-averaged wavelet spectra. The dashed curves correspond to the parameterization given by Equations~\ref{eq:wavelet_global_probability},~\ref{eq:time_avg_wavelet_para_a} and~\ref{eq:time_avg_wavelet_para_w}.}
	\label{fig:torrence_statistics_average_spectrum}
\end{figure}

The results are shown in Figure~\ref{fig:torrence_statistics_spectrum}. The histogram-style curves represent the Monte-Carlo-derived relations between $P_g(m)$ and $P(m)$ for $\delta j=1/8$ and varying values of $N$. The curves saturate at one, and this for smaller $P(m)$ as $N$ is large, reflecting the fact that the probability to have power greater than the threshold in at least one bin increases quickly with the number of bins. Almost identical curves are obtained for values of $\delta j$ smaller than 1/8, but they differ somewhat for larger values. TC98 mention that 1/2 is the largest $\delta j$ that still gives adequate sampling in scale for the Morlet wavelet. Our tests indicate that, at least from the point of view of the global confidence levels, $\delta j=1/8$ should be the maximum used.

We fitted the $\delta j\leq 1/8$ curves with Equation~\ref{eq:wavelet_global_probability} and found that they can be described 
by the following parameterization of the fitted $a$ and $n$ (dashed curves in Figure~\ref{fig:torrence_statistics_spectrum})
\begin{eqnarray}
a&=&0.810\ (N_{out}\delta j)^{0.011}\label{eq:wavelet_para_a}\\
n&=&0.491\ (N_{out}\delta j)^{0.926},\label{eq:wavelet_para_w}
\end{eqnarray}
\noindent
where $N_{out}$ is the number of bins outside the COI. The exponent 0.926 reflects the slightly slower increase of $n$ with $N_{out}\delta j$ compared to the expected linear relationship. The coefficient $a$ varies slowly with $N_{out}\delta j$, which corresponds to the fact that the curves are parallel to each other. Furthermore it is close to 1, which implies that it only introduces small perturbations to $P(m)$.

Inverting Equation~\ref{eq:wavelet_global_probability} and using the parameterization given by Equations~\ref{eq:wavelet_para_a} and~\ref{eq:wavelet_para_w}, one can now compute the {\it local} probability (or confidence level) that should be used to achieve a chosen {\it global} probability (or confidence level)
\begin{equation}
P(m)=\left(1 - \left(1 - P_g(m)\right)^{1/n}\right)^{1/a}.
\label{eq:inverted_wavelet_global_probability}
\end{equation}
\noindent
For example, assuming the $N=1024$ data points time series of Figure~\ref{fig:random_wavelet}, we have $a=0.892$ and $n=1595$ with the wavelet parameters used in this paper. The 95\% {\it global} confidence level ($P_g(m)=0.05$) thus corresponds to $P(m)=(1-(1-0.05)^{1/1595})^{1/0.892}=9\times 10^{-6}$, i.e. to a 99.999\% {\it local} confidence level. This value can in turn be used as input to the TC98 code (see the Appendix for practical details), and the structure visible in the bottom left panel of Figure~\ref{fig:random_wavelet} is now correctly rejected (no light blue contours).

\subsubsection{Global Time-averaged Confidence Levels}

Using the same Monte-Carlo simulations, we derived global confidence levels for time-averaged Morlet wavelet spectra. As shown by TC98, time-averaging over the $\chi_2^2$ distributed points of the wavelet spectrum results in the averaged power being distributed as $\sigma(\nu)\chi_d^2/d$, with $d$ being the number of degrees of freedom at each scale given by Equation~\ref{eq:time_averaged_dof}. For a given local confidence level, the corresponding threshold $m$ is thus a function of scale. But since all of the bins of the time-averaged spectrum have the same probability $P(m(s))$ to be above $m(s)\sigma(\nu)$ by chance, the global probability for at least one bin to be above $m(s)\sigma(\nu)$ should again follow a relation described by Equation~\ref{eq:wavelet_global_probability}, and we now expect $n$ to be proportional to $S_{out}\delta j$, the number of scales for which at least one bin is outside the COI, normalized by the resolution.

Figure~\ref{fig:torrence_statistics_average_spectrum} is the equivalent of Figure~\ref{fig:torrence_statistics_spectrum} for time-averaged spectra. As previously, the curves are all similar for $\delta j\leq 1/8$. The Monte-Carlo results  can be described by the following parameterization of $a$ and $n$ 
\begin{eqnarray}
a&=&0.805+0.45\times 2^{-S_{out}\delta j}\label{eq:time_avg_wavelet_para_a}\\
n&=&1.136\ (S_{out}\delta j)^{1.2}.\label{eq:time_avg_wavelet_para_w}
\end{eqnarray}
\noindent
Instead of the expected linear relationship, we had to raise $S_{out}\delta j$ to the power 1.2 in order to reproduce the fitted values of $n$. The dependence of $a$ with $S_{out}\delta j$ is not intuitive but it is always close to 1. 

As before, we can now use the above expressions to compute for a time-averaged spectrum the {\it local} confidence level that should be used to achieve a chosen {\it global} confidence level. For Figure~\ref{fig:random_wavelet}, we have $a=0.807$ and $n=12.7$. The 95\% {\it global} confidence level ($P_g(m)=0.05$) thus corresponds to $P(m)=(1 - (1-0.05)^{1/12.7})^{1/0.807}= 1\times 10^{-3}$, i.e. to a 99.89\% {\it local} confidence level. This value can in turn be used as input to the TC98 code.

We also ran Monte-Carlo simulations for the Paul wavelet and found results similar to those presented above for the Morlet wavelet: the coefficients $n$ are close to being proportional to the number of bins outside the COI and the coefficients $a$ are close to 1. We can use parameterizations of the same form as before, i.e. for the wavelet spectrum
\begin{eqnarray}
a&=&0.817\ (N_{out}\delta j)^{0.011}\label{eq:paul_para_a}\\
n&=&0.320\ (N_{out}\delta j)^{0.926},\label{eq:paul_para_w}
\end{eqnarray}
\noindent
and for the time-averaged spectrum
\begin{eqnarray}
a&=&1.02 + 0.70\times 1.42^{-S_{out}\delta j}\label{eq:time_avg_paul_para_a}\\
n&=&1.0 + 0.56\ (S_{out}\delta j)^{1.2}.\label{eq:time_avg_paul_para_w}
\end{eqnarray}


As a final note, it is important to realize that a peak of power may be above the {\it global} confidence level in the {\it time-averaged} wavelet spectrum  while the power  may be below the {\it global} confidence level at all times in the {\it wavelet} spectrum at the corresponding frequency. By analogy with the Fourier case, we derived {\it global} confidence levels to test individual peaks, not the probabilities of occurrence of extended structures. As shown in \S~\ref{sec:specificities}, this latter is simply given by the  probability associated with the power in the  time-averaged spectrum. Thus, while surpassing the {\it global} confidence level confirms the significance of a peak of power in the {\it wavelet} spectrum, the converse is not true. Before discarding as insignificant a peak in the {\it wavelet} spectrum that is below the {\it global} confidence level , one should check whether or not a corresponding peak is present and above the {\it global} confidence level in the {\it time-averaged} wavelet spectrum.
\section{Re-analysis of AIA time series\label{sec:aia_application}}

\begin{figure}[htbp!]
	\centering
		\includegraphics[width=0.456\textwidth]{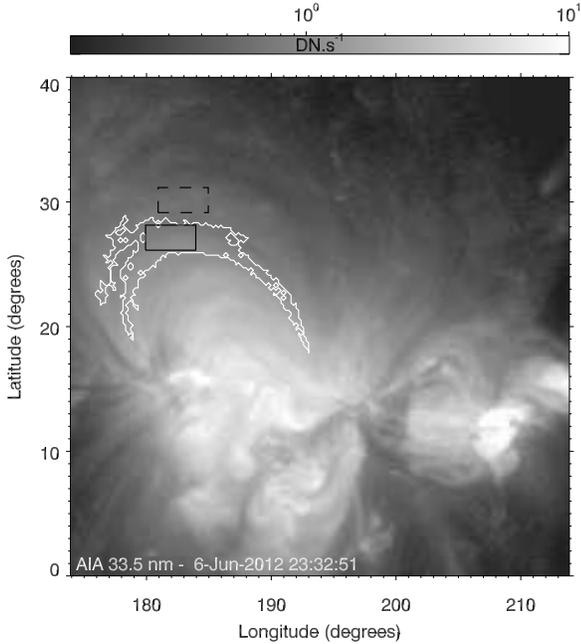}
		\caption{Middle frame of the 154 hr long sequence \sdoaia 33.5~nm sequence remapped to heliographic coordinates. \cite{Froment2015} detected excess Fourier power in the loop-shaped white contour and performed a detailed multi-wavelength analysis on the time series of intensities averaged over the black box. Section~\ref{sec:aia_application} describes the Fourier and wavelet re-analysis of this time series (see Figure~\ref{fig:case_1_335}). The dashed box delineates a nearby reference region (Figure~\ref{fig:case_1_335_reference}).}
	\label{fig:aia_mid_frames}
\end{figure}

\begin{figure*}[htbp!]
	\centering
		\includegraphics[width=\textwidth]{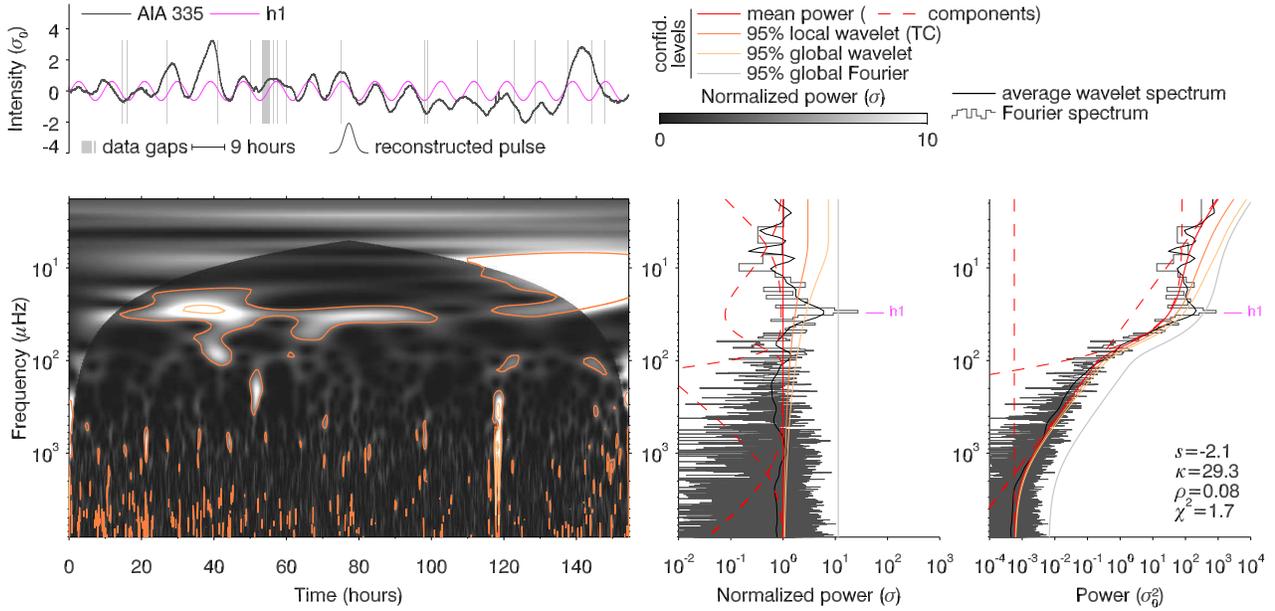}
		\caption{Fourier and wavelet analysis of the intensity time series averaged over the black box of Figure~\ref{fig:aia_mid_frames}. The bottom left panel shows the {\it whitened} wavelet spectrum (see~\S\ref{sec:aia_application}). The peak of Fourier power labeled h1 at 30~\muhz\ (9 hr) has a $1.7\times 10^{-8}$ probability of occurrence. The corresponding Fourier component is over-plotted on the time series in magenta. The equivalent peak in the time-averaged spectrum is also  well above the {\it global} confidence level (yellow curve). It corresponds to the elongated structure visible at the same frequency in the wavelet spectrum. Such a long-lived structure has a $7\times 10^{-11}$ probability of occurrence at this frequency. The vertical streak of wavelet power above $2\times 10^2$~\muhz\ around 118 hr is caused by the two small short-lived impulsive events visible in the time series.}
	\label{fig:case_1_335}
\end{figure*}

\begin{figure*}[htbp!]
	\centering
		\includegraphics[width=\textwidth]{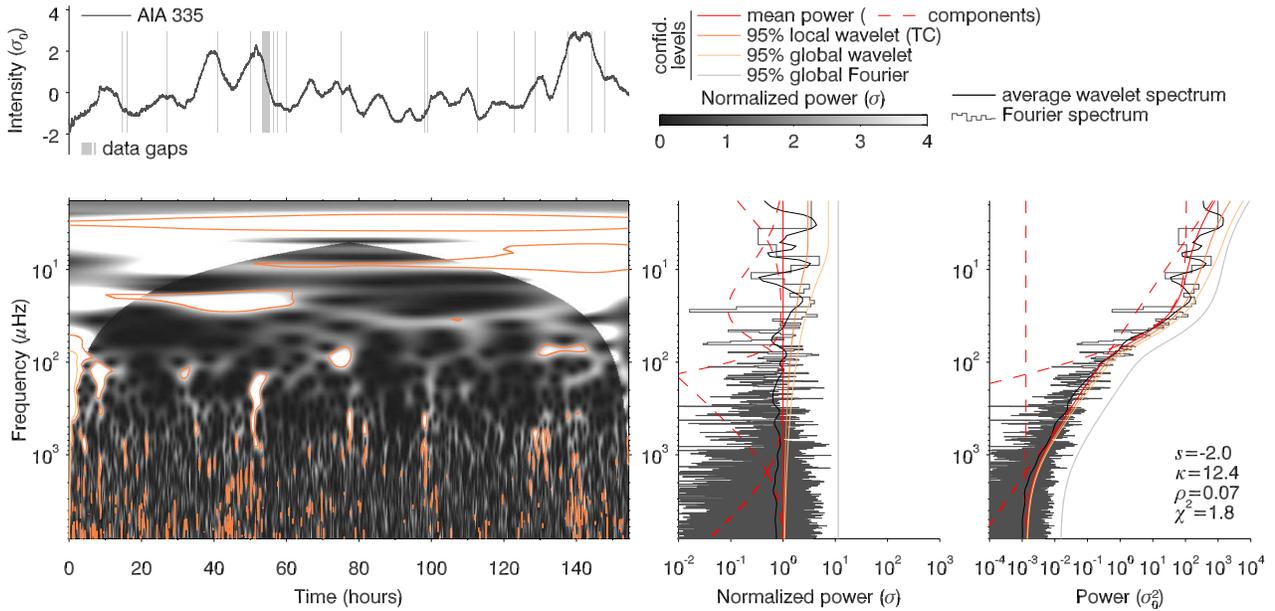}
		\caption{Same as Figure~\ref{fig:case_1_335} for the reference region of Figure~\ref{fig:aia_mid_frames} (dashed box). Taking into account all the degrees of freedom, no significant power surpasses the {\it global} confidence levels in the Fourier, time-averaged wavelet or wavelet spectra. To facilitate comparison, the power spectra use the same scaling.}
	\label{fig:case_1_335_reference}
\end{figure*}

We present a re-analysis of the 33.5~nm \sdoaia time series corresponding to one of the events studied by \citet{Froment2015}. Figure~\ref{fig:aia_mid_frames} shows the middle image of the one minute cadence, 9202 frame-long sequence starting 2012 June 3 at 18:00 UT and ending 2012 June 10 at 04:29 UT, remapped to the heliographic coordinates system used to compensate the solar differential rotation (see \cite{Auchere2005} for details on the projection method). The white contour delimits the region automatically detected in the outer part of NOAA AR 11499 by the algorithm described in \citet{Auchere2014}. It clearly delineates a bundle of loops whose length can estimated from magnetic field extrapolations to be about 280~Mm. The black box defines the area manually selected by \citet{Froment2015} for detailed analysis, which encloses the region where the maximum of power is observed. The dashed box delineates a nearby reference region of identical surface area chosen outside the white contour or excess power. The time series of intensities averaged over these boxes, normalized to their standard deviation $\sigma_0$, are shown in the top left panels of Figure~\ref{fig:case_1_335} and~\ref{fig:case_1_335_reference}, respectively. 
Data gaps, defined as the intervals during which no data exist within 30 s of an integer number of minutes since the beginning, represent 0.7\% of the sequence and are represented by the vertical gray bars, the height of which also represents the range of variation of the intensity.
The gaps have been filled with linear interpolations between the nearest data points. Since we used a one minute-cadence sample of the original 12~s cadence AIA data, the remainder of the time series was considered evenly spaced and thus kept as-is. The TC98 code zero-pads the times-series up to the next-higher power of two ($2^{14}$ in this case), while for Fourier analysis we apodized them using the Hann window.

In Figures~\ref{fig:case_1_335} and~\ref{fig:case_1_335_reference}, the right-hand panel gives the Fourier (histogram-style) and time-averaged wavelet (black solid line) spectra on a log-log scale. As discussed in \S~\ref{sec:detrending}, the power spectra of solar coronal time series generally exhibit a power-law dependence with frequency caused by a background of stochastic fluctuations. 
However, while the power spectra of Figures~\ref{fig:case_1_335} and~\ref{fig:case_1_335_reference} exhibit an overall power-law behavior, they depart significantly from a true power-law, as do many of the spectra studied by \citet{Auchere2014}.

Several effects indeed commonly produce complex spectral shapes. If different regimes of turbulence are present in the observed plasma, one can expect the spectrum to have one or several breaks due to the different power-law exponents in different frequency ranges, like in the solar wind \citep[e.g.][]{Bruno2013}. In addition, the superimposition on the background emission of one or a few sporadic transients also creates humps in the spectrum. In practice, the envelope of the power spectrum of many types of pulses can be modeled by a kappa function
\begin{equation}
\text{K}_\rho(\nu)=\left(1+\frac{\nu^2}{\kappa \rho^2}\right)^{-\frac{\kappa+1}{2}},
\label{eq:kappa_function}
\end{equation}
\noindent
$\rho$ being the width of the PSD and $\kappa$ defining the extent of its high-frequency wing. We thus chose to fit the Fourier spectrum with a background model of the form
\begin{equation}
\sigma(\nu)=A\nu^s+B\text{K}_\rho(\nu)+C.
\label{eq:kappa_model}
\end{equation}

\noindent
The first term represents the power-law dependence caused by the background of stochastic fluctuations present in most solar coronal time series. The second term is a kappa function that accounts for the possible presence of pulses in the time series. Finally, the constant $C$ corresponds to the high-frequency white noise component expected from photon statistics.

In the right-hand panels of Figures~\ref{fig:case_1_335} and~\ref{fig:case_1_335_reference}, the solid red curve shows the resulting models of mean power $\sigma(\nu)$, and the three dashed red curves correspond to the individual components. The two power spectra could be satisfactorily fitted with the same model, the reduced chi-squares being 1.7 and 1.8 respectively. In both cases, the white noise components dominate above 2~\muhz\ and the power-law slopes~$s$ are similar. The kappa function components create humps around 3~\muhz\ in both cases because they have similar widths, which is the signature of transients of similar duration in the two time series. 
At this stage, it is convenient (and justified from Equation~\ref{eq:fourier_wavelet_local_distribution}) to normalize the spectra to $\sigma(\nu)$ in order to better visualize possible deviations from the random $\chi_2^2$ distributed variations around the mean power. The middle panels display the same information as the right panels, but for the {\it whitened} power spectra.

Given the number of data points, the 95\% {\it global} Fourier confidence level is at 11.4~$\sigma$ (Equation~\ref{eq:fourier_global_probability}, gray lines). In Figure~\ref{fig:case_1_335}, the peak at 30~\muhz\ labeled h1 reaches 26.3~$\sigma$, which corresponds to a {\it global} probability of occurrence of $1.7\times10^{-8}$ (Equation~\ref{eq:fourier_global_probability}), while the Fourier power spectrum of the reference time series is everywhere below the 95\% {\it global} confidence level. At 5.8~$\sigma$, the time-averaged wavelet spectrum also peaks well above the corresponding 95\% {\it global} confidence level (yellow curve, 2.9~$\sigma$ at 30~\muhz\ from Equations~\ref{eq:time_averaged_dof} and ~\ref{eq:inverted_wavelet_global_probability}-\ref{eq:time_avg_wavelet_para_w}). The associated {\it global} probability is $6\times10^{-7}$, i.e. 35 times larger than that derived from the Fourier spectrum, yet still extremely small. The reference time-averaged wavelet power exceeds the 95\% {\it local} confidence level at 20~\muhz\ (orange curve), but this nearby peak is excluded at the 95\% {\it global} confidence level. The single Fourier component corresponding to the h1 peak is plotted in magenta in the top left panel of Figure~\ref{fig:case_1_335}. Comparison with the time series indicates that, except for the last pulse, the repetition period of 9 hr is very regular.

The bottom left panels of Figures~\ref{fig:case_1_335} and~\ref{fig:case_1_335_reference} show the {\it whitened} Morlet wavelet power spectra, i.e. normalized at each time step by the estimated mean Fourier power (red curves in the right-hand panels).\footnote{Note that by using the same background at each time step, one assumes the stationarity of the random  process against which the significance of the observed power is tested (TC98).} They are the counterparts of the whitened spectra of the middle panels. From Equation~\ref{eq:fourier_wavelet_local_distribution}, the 95\% {\it local} confidence level is at 3~$\sigma$ and, as expected given the large number of points in the spectrum, many points outside the COI lie above it for both time series (orange contours). The 95\% {\it global} confidence level is at 14~$\sigma$ (Equations~\ref{eq:fourier_wavelet_local_distribution} and~\ref{eq:wavelet_para_a}-\ref{eq:inverted_wavelet_global_probability}). In the reference wavelet spectrum, the power outside the COI does not exceed 8~$\sigma$. In contrast, in the bottom left panel Figure~\ref{fig:case_1_335}, the main peak of power, around 30~\muhz\ and 35~hours, reaches 16.3~$\sigma$. This peak has a $4.7\times 10^{-3}$ {\it global} probability to occur by chance (Equations~\ref{eq:wavelet_global_probability} and~\ref{eq:fourier_wavelet_local_distribution}). Around this frequency, the power remains above the 95\% {\it local} confidence level during most of the time series, except between 100 and 115 hr. Using Equation~\ref{eq:binomial}, the probability of occurrence of a structure of this length at a given frequency is $7.3\times 10^{-11}$, which is comparable to the {\it local} probability associated with the corresponding peak in the time-averaged spectrum, as expected from \S~\ref{sec:specificities} and Figure~\ref{fig:binomial_correlation}.

\section{SUMMARY\label{sec:summary}}

The wavelet code described in TC98 is widely used in a variety of scientific fields, as indicated by the 1741 citations of the paper referenced in the Astrophysics Data System as of 2016 March 21. It is used both for the analysis of observations~\citep[e.g.,][]{Ireland1999, DeMoortel2000, Williams2001, DePontieu2010, McIntosh2010, Madsen2015} and of numerical simulations~\citep[e.g.,][]{Nakariakov2004, Meszarosova2014, Pascoe2014}. The popularity of this code is due to its ease of use, its ability to produce clear and convincing graphics, as well as its output of rigorous quantitative confidence levels. However, it should not be used as a black box, for the confidence levels are linked to background models and, as we have demonstrated in \S~\ref{sec:noise_model}, the white and red noise models built into the code generally cannot represent the power spectra of solar coronal time series. The problem is potentially worse if detrending is applied to the time series, for this pre-processing distorts its power spectrum (section~\ref{sec:detrending}). 
In addition, even assuming an adequate background model, the confidence levels are {\it local}, i.e. they do not take into account the total number of degrees of freedom in the wavelet and time-averaged wavelet spectra. In most cases, it is thus likely that at least one bin of the spectrum lies above the TC98 confidence levels. Both effects -- improper background model and {\it local} confidence levels -- are prone to produce false positives.

These limitations of the TC98 code can nevertheless be easily overcome. First, a noise model suited to the considered data set must be found. We propose in \S~\ref{sec:aia_application} a function (Equation~\ref{eq:kappa_model}) that satisfactorily fits the power spectra of many time series, but the adequacy of the chosen model should always be verified. Generalizing the principle introduced by \cite{Scargle1982}, {\it global} confidence levels taking into account the total number of degrees of freedom can then be computed for the Fourier, wavelet and time-averaged wavelet spectra, as described in \S~\ref{sec:confidence_levels}. The Appendix describes the practical details. Note that the empirically derived coefficients are valid only for the Morlet and Paul wavelets and for time series of up to $2^{14}$ samples. Other Monte-Carlo simulations should be carried out to determine their equivalents for other wavelets and/or longer series.

Following this methodology, we re-analyzed in \S~\ref{sec:aia_application} one of the \sdoaia intensity time series in which \cite{Froment2015} detected 9 hr period pulsations. The Fourier and time-averaged wavelet spectra both exhibit a strong peak at 30~\muhz\ with associated {\it global} probabilities of occurrence below $10^{-6}$. The corresponding structure in the wavelet spectrum lasts for most of the sequence and also peaks above the 95\% {\it global} confidence level. In contrast, no significant power could be detected in a nearby reference region, which implies a sharp spatial boundary of the detected periodic phenomenon, as was already visible in the power maps (Figure~4) of  \cite{Froment2015}.

The present analysis of the Fourier and wavelet confidence levels, combined with our previous investigation of potential instrumental and geometrical artefacts \citep{Auchere2014}, lead us to conclude beyond reasonable doubt that the detected pulsations are of solar origin.

\begin{acknowledgements}
The authors would like to thank John Leibacher for relentlessly trying to demonstrate the instrumental origin of the pulsation phenomena reported here. The authors acknowledge the use of the wavelet code by \cite{Torrence1998}. The authors acknowledge the use of \sdoaia data. This work used data provided by the MEDOC data and operations centre (CNES/CNRS/Univ. Paris-Sud), http://medoc.ias.u-psud.fr/. Following the design principles advocated by \citet{Tufte2001}, we aimed to maximize the data-ink ratio of our graphics.
\end{acknowledgements}

\bibliographystyle{apj}
\bibliography{bibliography}

\appendix

%
%

\section{HOW TO USE THE TC98 CODE WITH CUSTOM NOISE MODELS AND CONFIDENCE LEVELS\label{sec:tc98_practical}}

{\color{red}Typos in the original version of the paper (as published in ApJ) prevented the code snippets provided in the Annex from running properly. These typos have been corrected in the present version of the paper. In addition, a full demonstration code is available at \url{https://idoc.ias.u-psud.fr/MEDOC/wavelets\_tc98}}\\

Here we provide practical details on how to use the TC98 code with any noise model (e.g. the one described in \S~\ref{sec:aia_application}, Equation~\ref{eq:kappa_model}) instead of the built-in white or red noises, and with the {\it global} confidence levels introduced in \S~\ref{sec:confidence_levels}.
Let {\tt data} be an {\tt n} element long 1D floating point array containing a time series of step {\tt dt}. The Morlet wavelet power spectrum is obtained with the following Interactive Data Language commands
\begin{Verbatim}[commandchars=\\\{\}]
mother = 'Morlet'
s0 = 2*dt
dj = 1/8.0
j1 = FIX(alog(n/2.0)/alog(2)/dj)
wave = WAVELET(data,dt,PERIOD=period,S0=s0,PAD=1,DJ=dj,$
                                J=j1,MOTHER=mother,COI=coi,SCALE=scale)
power = ABS(wave)^2
\end{Verbatim}

Let us now assume that we have fitted the Fourier power spectrum of the time series with, e.g., Equation~\ref{eq:kappa_model} and that the result as a function of the {\tt period} values returned by the above call to the {\tt WAVELET} function is stored in the 1D floating point array {\tt background\_fit\_period}.
The trick to compute the corresponding confidence levels is to exploit the {\tt GWS} keyword that is normally provided in the code to use the time-averaged wavelet spectrum as a background noise model. For a {\it local} confidence level of 95\% we use

\begin{verbatim}
local_siglvl = 0.95
local_signif = WAVE_SIGNIF(data,dt,scale,0, $
                           SIGLVL=local_siglvl,GWS=background_fit_period,MOTHER=mother) 
local_signif = REBIN(TRANSPOSE(local_signif),n,j1+1)                          
\end{verbatim}

\noindent
The last line replicates the {\tt local\_signif} 1D array at every time step to form an array of the same size as {\tt power}. The points of the wavelet spectrum above the 95\% {\it local} confidence level then correspond to the elements of the {\tt power} array greater than {\tt local\_signif}. Now for a {\it global} confidence level of 95\%, we first use Equations~\ref{eq:wavelet_para_a}-\ref{eq:inverted_wavelet_global_probability}, to compute the corresponding {\it local} confidence level

\begin{Verbatim}[commandchars=\\\{\}]
global_siglvl = 0.95
jcoi = ALOG(coi/1.033/s0)/ALOG(2)/dj
Nout = TOTAL(jcoi>0)
a_coeff = 0.810*(Nout*dj)^0.011
n_coeff = 0.491*(Nout*dj)^0.926
local_siglvl = 1 - (1 - global_siglvl^(1/n_coeff))^(1/a_coeff)
\end{Verbatim}

\noindent
Note that we estimate {\tt Nout} -- the number of bins of the spectrum outside the COI -- by using the {\tt coi} array returned by the {\tt WAVELET} function, with 1.033 being the Fourier factor for the Morlet wavelet (see TC98). Then we use the same syntax as above to call the {\tt WAVE\_SIGNIF} function, the only difference being the different value of the {\tt local\_siglvl} variable
\begin{verbatim}
global_signif = WAVE_SIGNIF(data,dt,scale,0, $
                            SIGLVL=local_siglvl,GWS=background_fit_period,MOTHER=mother)
global_signif = REBIN(TRANSPOSE(global_signif),n,j1+1)                                                 
\end{verbatim}

\noindent
The points of the wavelet spectrum above the 95\% {\it global} confidence level then correspond to the elements of the {\tt power} array greater than {\tt global\_signif}. Similarly, we can compute the 95\% {\it global} confidence level for the time-averaged wavelet spectrum after using Equations~\ref{eq:inverted_wavelet_global_probability},~\ref{eq:time_avg_wavelet_para_a} and~\ref{eq:time_avg_wavelet_para_w} to compute the corresponding {\it local} significance level 
\begin{Verbatim}[commandchars=\\\{\}]
Sout = MAX(jcoi)
a_scl_coeff = 0.805+0.45*2^(-Sout*dj)
n_scl_coeff = 1.136*(Sout*dj)^1.2
time_avg_local_siglvl = 1 - (1 - global_siglvl^(1/n_scl_coeff))^(1/a_scl_coeff)
dof = n - scale/dt
time_avg_global_signif = WAVE_SIGNIF(data,dt,scale,1, $
                                     SIGLVL=time_avg_local_siglvl, $                           
                                     GWS=background_fit_period,DOF=dof,MOTHER=mother)                           
\end{Verbatim}
\noindent 
The {\tt WAVE\_SIGNIF} function is now called with setting the {\tt sigtest} argument to 1 and setting the {\tt DOF} keyword to the number of data points minus half the number of those in the COI. The resulting array has the same number of elements as the time averaged spectrum and is used to determine which bins are above the 95\% {\it global} confidence level.

\end{document}